\documentclass[12pt,twoside]{article}
\usepackage{enumerate}         
\usepackage{amsmath}  
\usepackage{amssymb}
\usepackage{fancyhdr}
\usepackage{bm} 
\usepackage{graphicx,color}
\numberwithin{equation}{section}
\begin{document}
\begin{center}
\Large
{\bf AN ENTROPIC PATHWAY TO MULTIVARIATE \\GAUSSIAN DENSITY }
\end{center}
\begin{center}
{\bf H.J. Haubold}\footnote[1]{Corresponding author:
\\
Email address: HANS.HAUBOLD@UNVIENNA.ORG}, {\bf  A.M. Mathai}$^2$, {\bf  S. Thomas}$^3$
\vskip.3cm
{\footnotesize
$^1$Office for Outer Space Affairs, United Nations, Vienna International Centre, P.O. Box 500, A-1400, Vienna, Austria.\\
$^2$Centre for Mathematical Sciences Pala Campus, Arunapuram P.O., Pala-686 574, Kerala, India and Department of Mathematics and Statistics, McGill University, Montreal, Canada H3A 2K6.\\
$^3$Department of Statistics, St. Thomas College, Palai, Arunapuram P.O., Pala-686 574, Kerala, India}
\end{center}
\vskip.5cm
\noindent
{\bf Abstract}

A general principle called ``conservation of the ellipsoid of concentration" is introduced and a generalized entropic form of order $\alpha$ is optimized under this principle. It is shown that this can produce a density which can act as a pathway to multivariate Gaussian density. The resulting entropic pathway contains as special cases the Boltzmann-Gibbs (Shannon) and Tsallis (Havrda-Charv$\acute{\rm a}$t) entropic forms.

\section{Introduction}

The normal (Gaussian) distribution is a family of continuous probability distributions and is ubiquitous in the field of statistics and probability (Feller [4]). The importance of the normal distribution as a model of quantitative phenomena is due to the central limit theorem. The normal distribution maximizes Shannon entropy among all distributions with known mean and variance and in information theory, Shannon entropy is the measure of uncertainty associated with a random variable.

In statistical mechanics, Gaussian (Maxwell-Boltzmann) distribution maximizes the Boltzmann-Gibbs entropy under appropriate constraints (Gell-Mann and Tsallis [7]). Given a probability distribution $P = \left\{{p_i}\right\}\;\; (i = 1,..., N)$, with $p_i$ representing the probability of the system to be in the {\it ith} microstate, the Boltzmann-Gibbs entropy is $S(P) = - k \sum_{i=1}^ N p_i ln p_i$, where $k$ is the Boltzmann constant and $N$ the total number of microstates. If all states are equally probable it leads to the Boltzmann principle $S = k\; ln W\;\; (N=W)$. Boltzmann-Gibbs entropy is equivalent to Shannon's entropy if $k=1$. 

A generalization of Boltzmann-Gibbs extensive statistical mechanics is known as Tsallis non-extensive statistical mechanics (Swinney and Tsallis [5], Abe and Okamoto [6]). Tsallis discovered the generalization of Shannon's entropy to non-extensivity as $S(P,q) = (\sum_{i=1}^ N p_i^q - 1)/(1-q)$. For $q\rightarrow 1$, Shannon's entropy is recovered. Tsallis introduced $q$-probabilities accommodating the fact that non-extensive systems are better described by power law distributions, $p_i^q$, now called $q$-probabilities. The $p_i^q$ are scaled probabilities where $q$ is a real parameter.

This paper, in Section 2, introduces a general principle called conservation of the ellipsoid of concentration and maximizes a generalized entropic form of order $\alpha$, containing Shannon (Boltzmann-Gibbs), R$\acute{\rm e}$nyi, Havrda-Charv$\acute{\rm a}$t (Tsallis) entropies as special cases, under this principle, in Section 3. Normalizing constants are derived in Section 3.1 and mean value and covariance matrix in Section 3.2 for the cases $\alpha < 1$, $\alpha > 1$, and $\alpha = 1$. The pathway, characterized by $\alpha$ is shown to produce multivariate type-1 beta, Gaussian, and type-2 beta densities, respectively. In Section 3.3 a graphical representation of the pathway surface is shown. Section 4 draws conclusions.

\section{Conservation of the ellipsoid of concentration}

Consider a $q\times 1$ vector $X, X'=(x_1,\ldots, x_q)$, where a prime denotes the transpose. The components $x_1,\ldots, x_q$ may be real scalar mathematical variables or random variables describing various components in a physical system. Each component in $X$ can be assumed to have a finite mean value and variance. If $E$ denotes the expected value, the value on the average in the long-run, then we can assume $E(x_i)=\mu_i<\infty$ for $i=1,\ldots, q$. Let $\mu'=(\mu_1,\ldots,\mu_q)$. Similarly one can assume the expected dispersion in each component to be finite. The square of a measure of dispersion is given by the variance or Var$(x_i)$. That is, Var$(x_i)<\infty$. The components may be correlated or may have pair-wise joint variations. A measure of pair-wise joint variation is covariance between $x_i$ and $x_j$ or Cov$(x_i,x_j)=E[x_i-E(x_i)][x_j-E(x_j)]=v_{ij}$ so that when $i=j$ we have Var$(x_i)=v_{ii}$. The matrix of such variances and covariances is the covariance matrix in $X$, denoted by Cov$(X)=E(X-E(X))(X-E(X))'=V=(v_{ij})$.  Note that $V$ is real symmetric when $x_i,i=1,\ldots,q$ are real, and $V$ is at least non-negative definite. Let us assume that no component in the $q\times 1$ vector $X$ is a linear function of other components so that we can take $V$ to be nonsingular. This will then imply that $V$ is positive definite. That is, $V=V'>0$. Let $V^{\frac{1}{2}}$ be the positive definite square root of the positive definite matrix $V$.

Standardization of a component $x_i$ is achieved by relocating it at $\mu_j$ and by rescaling it by taking $y_i=\frac{x_i-\mu_i}{\sqrt{Var(x_i)}}$ so that $E(y_i)=0$ and Var$(y_i)=1$. Similarly, standardization of the $q\times 1$ vector $X$ is achieved by a linear transformation on $X-\mu$, namely, $Y=V^{-\frac{1}{2}}(X-\mu)$ so that $E(Y)=O$ and Cov$(Y)=I$ where $I$ is the identity matrix. The Euclidean norm in $Y$ is then $[Y'Y]^{\frac{1}{2}}=[(X-\mu)'V^{-1}(X-\mu)]^{\frac{1}{2}}$. This scalar quantity $(X-\mu)'V^{-1}(X-\mu)$ has many interpretations in different disciplines. A measure of distance between $X$ and $\mu$ is any norm $||X-\mu||$. But if we want to accommodate the joint variations in the components $x_1,\ldots, x_q$ as well as the fact that the variances of the components may be different then we consider a generalized distance between $X$ and $\mu$. One such square of the generalized distance is the square of the Euclidian norm in $Y$ or $Y'Y=(X-\mu)'V^{-1}(X-\mu)$. For a given constant $c>0, (X-\mu)'V^{-1}(X-\mu)=c$ defines the surface of an ellipsoid since $V$ is positive definite. This ellipsoid is known as the ellipsoid of concentration of $X$ around its expected value $\mu$. If we assume that $c$ is fixed, for example $c=1$ which implies  $(X-\mu)'V^{-1}(X-\mu)=1$ then this assumption is equivalent to saying that the standardized $X$, namely, $Y$ is a point on the surface of a hypersphere of radius 1. When it is assumed that the ellipsoid of concentration is a fixed finite quantity what we are saying is that the generalized distance of $X$ from $\mu$ is fixed and finite. This is the principle of conservation of the ellipsoid of concentration.

\section{Generalized entropic form of order $\alpha$}

Let $f(X)$ be a real-valued scalar function of $X$ where $X$ could be a scalar quantity or a $q\times 1$ vector, $q>1$, or $p\times q$ matrix, $p>1,q>1$. Let us assume that the elements in $X$ are real scalar random variables. Then $f(X)$ can define a density provided $\int_Xf(X){\rm d}X=1$ and $f(X)\geq 0$ for all $X$. If $\int_Xf(X){\rm d}X=h<\infty$ then $g(X)=\frac{1}{h}f(X)$ is a density provided $f(X)\geq0$ for all $X$. Here ${\rm d }X$ denotes the wedge product of the differentials in $X$. For example, ${\rm d }X={\rm d} x_{11}\wedge {\rm d }x_{12}\wedge\ldots \wedge{\rm d}x_{1q}\wedge {\rm d}x_{21} \wedge \ldots \wedge{\rm d}x_{pq}$ if $X$ is $p\times q$ and all elements in $X$ are functionally independent. A measure of uncertainty or information in $X$ or in $f(X)$ is measured by Shannon entropy defined by 
\begin{equation}
S(f)=-\int_X f(X)\ln f(X){\rm d} X
\end{equation}
when $f$ is continuous, where $X$ may be scalar or vector or a general matrix and $f$ is the density of $X$. There are generalizations of $S(f)$, some of them are listed in Mathai and Rathie [1]. Some of these are the following (Mathai and Haubold [2]):\\
R$\acute{\rm e}$nyi's entropy $R_\alpha(f)=\displaystyle \frac{\ln [\int_X\{f(X)\}^\alpha{\rm d }X]}{1-\alpha},~\alpha\neq 1, \alpha>0$\\
Havrda-Charv$\acute{\rm a}$t entropy $H_\alpha(f)=\displaystyle\frac{\int_X[f(X)]^\alpha]{\rm d}X-1}{2^{1-\alpha}-1},~\alpha\neq 1, \alpha>0$\\
Tsallis' non-extensive entropy $T_\alpha(f)=\displaystyle \frac{\int_X[f(X)]^\alpha{\rm d}X-1}{1-\alpha},~\alpha \neq1,\alpha>0$\\
Non-extensive generalized entropic form $M_\alpha(f)=\displaystyle \frac{\int_X[f(X)]^{2-\alpha}{\rm d}X-1}{\alpha-1},~\alpha \neq1,\alpha<2$\\
Extensive generalized entropic form $M_\alpha^*(f)=\displaystyle \frac{\ln[f_X\{f(X)\}^{2-\alpha}{\rm d} X]}{\alpha-1},~\alpha \neq1,\alpha<2.$

Let us look into the problem of optimizing the non-extensive generalized entropic form $M_\alpha(f)$ under the principle of the conservation of the ellipsoid of concentration. That is, to optimize $M_\alpha(f)$ over all functional $f$, subject to the conditions
\[\mbox{ (i) }\int_X f(X){\rm d} X=1;~ \mbox{ (ii) } \int_X (X-\mu)'V^{-1}(X-\mu)f(X){\rm d} X=\mbox{ constant }\]
for all $f\geq 0$ for all $X$. If we apply calculus of variation technique then the Euler equation becomes 
\[\frac{\partial}{\partial f}\left[f^{2-\alpha}-\lambda_1 f+\lambda_2(X-\mu)'V^{-1}(X-\mu)f\right]=0,~\alpha<2\]
where $\lambda_1$ and $\lambda_2$ are Lagrangian multipliers, observing the fact that since $\alpha$ is fixed, optimization of $\frac{f^{2-\alpha}}{\alpha-1}$ is equivalent to optimizing $f^{2-\alpha}$ over all functional $f$. That is, 
\[f^{1-\alpha}=\frac{\lambda_1}{2-\alpha}\left[1-\frac{\lambda_2}{\lambda_1}(X-\mu)'V^{-1}(X-\mu)\right].\]
Either by taking $\frac{\lambda_2}{\lambda_1}=a(1-\alpha),~a>0$ or by taking the second condition as the expected value of $(1-\alpha)(X-\mu)'V^{-1}(X-\mu)$ is 1 where $1-\alpha$ denotes the strength of information in $f(X)$, see Mathai and Haubold [2], we have
\begin{equation}
f=\lambda[1-a(1-\alpha)(X-\mu)'V^{-1}(X-\mu)]^{\frac{1}{1-\alpha}}
\end{equation}
where $\lambda$ is the normalizing constant, $1-a(1-\alpha)(X-\mu)'V^{-1}(X-\mu)>0$. Observe that when $\alpha<1$ the form in (3.2) is that of a multivariate type-1 beta type density. When $\alpha>1$, writing $1-\alpha=-(\alpha-1)$ we have
\begin{equation}
f=\lambda[1+a(\alpha-1)(X-\mu)'V^{-1}(X-\mu)]^{-\frac{1}{\alpha-1}},~\alpha>1,~a>0.
\end{equation}
Note that (3.3) is a multivariate type-2 beta type density. But when $\alpha\to 1$ in (3.2) and (3.3) we have the form
\begin{equation}
f=\lambda {\rm e}^{-a(X-\mu)'V^{-1}(X-\mu)}.
\end{equation}
Note that $\lambda$ in (3.2), (3.3) and (3.4) are different, which are to be evaluated separately  for the three cases  of $\alpha<1,\alpha>1$ and $\alpha\to 1$. Thus (3.2) and (3.3) provide a pathway to the multivariate Gaussian density in (3.4). When $a=\frac{1}{2}$ the normalizing constant in (3.4) is 

\begin{eqnarray}
\nonumber
\lambda&=&\frac{a^{\frac{q}{2}}}{(\pi)^{\frac{q}{2}}|V|^{\frac{1}{2}}}\\
&=&\frac{1}{(2\pi)^{\frac{q}{2}}|V|^{\frac{1}{2}}}\mbox{ for }  a=\frac{1}{2}
\end{eqnarray}
or when $\alpha\to 1$ in (3.2) and (3.3).
 
\subsection{The normalizing constant $\lambda$}

Let us consider the case $\alpha<1$ first. Since the total integral is 1 we have
\begin{eqnarray*}
1&=&\int_Xf(X){\rm d}X\\
&=&\lambda|V|^{\frac{1}{2}}\int_Y[1-a(1-\alpha)(y_1^2+\cdots+y_q^2)]^{\frac{1}{1-\alpha}}{\rm d}Y,~Y=V^{-\frac{1}{2}}(X-\mu)\Rightarrow {\rm d}X=|V|^{\frac{1}{2}}{\rm d}Y.\\
&=&\lambda 2^q|V|^{\frac{1}{2}}\int\ldots \int_{y_j>0,j=1\ldots q, 1-a(1-\alpha)(y_1^2+\cdots+y_q^2)>0}[1-a(1-\alpha)(y_1^2+\cdots+y_q^2)]^{\frac{1}{1-\alpha}}{\rm d}Y.
\end{eqnarray*}
Put $u_j=a(1-\alpha)y_j^2\Rightarrow {\rm d}y_j=\frac{1}{2}\frac{u_j^{\frac{1}{2}-1}{\rm d}u_j'}{[a(1-\alpha)]^{\frac{1}{2}}},~\alpha<1$. Then 
\begin{eqnarray*}
1&=&\frac{\lambda|V|^{\frac{1}{2}}}{[a(1-\alpha)]^{\frac{q}{2}}}\int\ldots\int_{1-u_1-\cdots-u_q>0,0<u_j<1, j=1,\ldots,q} u_1^{\frac{1}{2}-1}\ldots u_q^{\frac{1}{2}-1}\\
&&\times (1-u_1-\cdots- u_q)^{\frac{1}{1-\alpha}}{\rm d}u_1\wedge\ldots\wedge {\rm d} u_q\\
&=&\frac{\lambda|V|^{\frac{1}{2}}}{[a(1-\alpha)]^{\frac{q}{2}}}\frac{\left[\Gamma\left(\frac{1}{2}\right)\right]^q\Gamma \left(\frac{1}{1-\alpha}+1\right)}{\Gamma\left(\frac{1}{1-\alpha}+1+\frac{q}{2}\right)}.
\end{eqnarray*}
by evaluating the integral with the help of a type-1 Dirichlet integral (Mathai [3]). Thus 
\begin{equation}
\lambda =\frac{\Gamma\left(\frac{1}{1-\alpha}+1+\frac{q}{2}\right)[a(1-\alpha)]^{\frac{1}{2}}}
{\Gamma\left(\frac{1}{1-\alpha}+1\right)|V|^{\frac{1}{2}}\pi^{\frac{q}{2}}}
\mbox{ for } \alpha<1.
\end{equation}
For $\alpha>1$, writing $1-\alpha=-(\alpha-1)$ and proceeding as above and then finally evaluating the integral with the help of a type-2 Dirichlet integral [Mathai [3]) we have
\begin{equation}
\lambda=\frac{[a(\alpha-1)]^{\frac{q}{2}}\Gamma\left(\frac{1}{\alpha-1}\right)}
{|V|^{\frac{1}{2}}\pi ^{\frac{q}{2}}\Gamma \left(\frac{1}{\alpha-1}-\frac{q}{2}\right)},~\frac{1}{\alpha-1}-\frac{q}{2}>0,~\alpha>1.
\end{equation}
When $\alpha\to 1$ do (3.6) and (3.7) go to (3.3)? This can be checked with the help of Stirling's formula which states that for $|z|\to \infty$ and $\varepsilon$ a bounded quantity,
\begin{equation}
\Gamma (z+\varepsilon)\approx \sqrt{2\pi}z^{z+\varepsilon-\frac{1}{2}}{\rm e}^{-z}.
\end{equation}
Note that for $\alpha<1$ and when $\alpha\to 1,~\frac{1}{1-\alpha}\to \infty$. Then applying Stirling's formula to $\Gamma\left(\frac{1}{1-\alpha}+1+\frac{q}{2}\right)$ and $\Gamma \left(\frac{1}{1-\alpha}+1\right)$ in (3.6) we have
\begin{eqnarray*}
\lambda&\to&\frac{\sqrt{2\pi}\left(\frac{1}{1-\alpha}+1+\frac{q}{2}\right)^{\frac{1}{1-\alpha}+1+\frac{q}{2}-\frac{1}{2}}{\rm e}^{-\frac{1}{1-a}}[a(1-\alpha)]^{\frac{q}{2}}}
{\sqrt{2\pi}\left(\frac{1}{1-\alpha}+1\right)^{\frac{1}{1-\alpha}+1-\frac{1}{2}}{\rm e}^{-\frac{1}{1-\alpha}}|V|^{\frac{1}{2}}\pi ^{\frac{q}{2}}}\\
&=&\frac{a^{\frac{q}{2}}}{\pi ^{\frac{q}{2}}|V|^{\frac{1}{2}}}
\end{eqnarray*}
which is the value of $\lambda$ in (3.5). Then when $\alpha$ approaches 1 from the left, (3.6) goes to (3.5). Similarly we can see that (3.7) also goes to (3.5) when $\alpha\to 1$ from the right. This constitutes the pathway to multivariate Gaussian density.

\subsection{The mean value and covariance matrix of $X$ in (3.2)}

\begin{eqnarray*}
E(X)&=&\int_XXf(X){\rm d}X=\mu\int_Xf(X){\rm d}X+\int_X(X-\mu)f(X){\rm d}X\\
&=&\mu+\lambda |V|^{\frac{1}{2}}\{V^{\frac{1}{2}}\int_YY[1-a(1-\alpha)Y'Y]^{\frac{1}{1-\alpha}}{\rm d}Y\},
\end{eqnarray*}
since $\int_Xf(X){\rm d}X=1$ and since $X-\mu=V^{\frac{1}{2}}Y$ when $Y=V^{-\frac{1}{2}}(X-\mu)\Rightarrow {\rm d}X=|V|^{\frac{1}{2}}{\rm d}Y.$ But $Y[1-a(1-\alpha)Y'Y]^{\frac{1}{1-\alpha}}$ is an odd function and hence the integral over $Y$ is null. Hence $E(X )=\mu$.
\begin{eqnarray*}
\mbox{Cov}(X)&=&E(X-E(X))(X-E(X))'\\
&=&E(X-\mu)(X-\mu)'\\
&=&V^{\frac{1}{2}}\{E(YY')\}V^{\frac{1}{2}}, Y=V^{-\frac{1}{2}}(X-\mu)\\
&=&\lambda|V|^{\frac{1}{2}}V^{\frac{1}{2}}\{\int_YYY'[1-a(1-\alpha)Y'Y]^{\frac{1}{1-\alpha}}{\rm d}Y\}V^{\frac{1}{2}}.
\end{eqnarray*}
Note that $YY'$ is a $q\times q$ matrix where the $(i,j)$th element is $y_iy_j$. For $i\neq j$ the integral over $Y$ is zero since $y_iy_j[1-a(1-\alpha)Y'Y]^{\frac{1}{1-\alpha}}$ is an odd function in $y_i$ as well as in $y_j$. The diagonal elements of $YY'$ are $y_1^2,\ldots,y_q^2$. The integral over one of them will be of the form 
\begin{eqnarray*}
\noalign{$\int_Yy_1^2[1-a(1-\alpha)Y'Y]^{\frac{1}{1-\alpha}}{\rm d}Y\mbox{ for } 1-a(1-\alpha)Y'Y>0 \mbox{ when } \alpha<1.$}
&=&2^q\int\ldots \int y_1^2(1-a(1-\alpha)Y'Y]^{\frac{1}{1-\alpha}}{\rm d}Y\mbox{ for } y_i>0, j=1\ldots q, ~\alpha<1 \mbox{ and } \\
&&1-a(1-\alpha) (y_1^2+\cdots+y_q^2)>0\\
&=&\frac{1}{[a(1-\alpha)]^{\frac{q}{2}+1}}\int\ldots\int u_1^{\frac{3}{2}-1}u_2^{\frac{1}{2}-1}\ldots
u_q^{\frac{1}{2}-1}(1-u_1-\cdots -u_q)^{\frac{1}{1-\alpha}} {\rm d}u_1\wedge\ldots \wedge {\rm d}u_q\\
&=&\frac{\frac{1}{2}\left[\Gamma\left(\frac{1}{2}\right)\right]^q\Gamma\left(\frac{1}{1-\alpha}+1\right)}
{[a(1-\alpha)]^{\frac{q}{2}+1}\Gamma\left(\frac{1}{1-\alpha}+1+\frac{q}{2}+1\right)},
\end{eqnarray*}
by using a type-1 Dirichlet integral. Now, substitute in (3.2) and (3.6) we have
\begin{equation}
\mbox{Cov}(X)=\frac{1}{2a(1-\alpha)\left[\frac{1}{1-\alpha}+1+\frac{q}{2}\right]}V=\frac{1}{2a[1+(1-\alpha)(1+\frac{q}{2})]} V,~\alpha<1.
\end{equation}
Observe that it is an interesting result because the covariance matrix in $X$ is not the parameter matrix $V$ in the model (3.2) and (3.6). For $\alpha>1$, proceeding as before, one has 
\begin{eqnarray}
\mbox{Cov}(X)&=&\frac{1}{2a[1-(\alpha-1)\left(\frac{q}{2}+1\right)}V,
\end{eqnarray}
for $\alpha>1, 1-(\alpha-1)\left(\frac{q}{2}+1\right)>0$ which implies $ 1<\alpha<1+\frac{1}{\frac{q}{2}+1}.$
Observe that when $a=\frac{1}{2}$ and $\alpha\to 1$ then (3.8) and (3.9) give the covariance matrix as $V$ which agrees with the multivariate Gaussian density. Hence the pathway for the covariance matrix is given in (3.8) and (3.9). 

\subsection{The pathway surface}

Let us look into the pathway model for the standard case. That is, for $\alpha<1$,
\[g_1(Y)=\frac{[a(1-\alpha)]^{\frac{q}{2}}\Gamma \left(\frac{1}{1-\alpha}+1+\frac{q}{2}\right)}
{\Gamma\left(\frac{1}{1-\alpha}+1\right)\pi ^{\frac{q}{2}}} [1-a(1-\alpha)(y_1^2+\cdots+y_q^2)]^{\frac{1}{1-\alpha}},\alpha<1\]
$1-a(1-\alpha)(y_1^2+\cdots+y_q^2)>0$. This is plotted for $q=2,a=1$ and for $\alpha=-0.5,0,0.5.$\\
\begin{center} 
\includegraphics{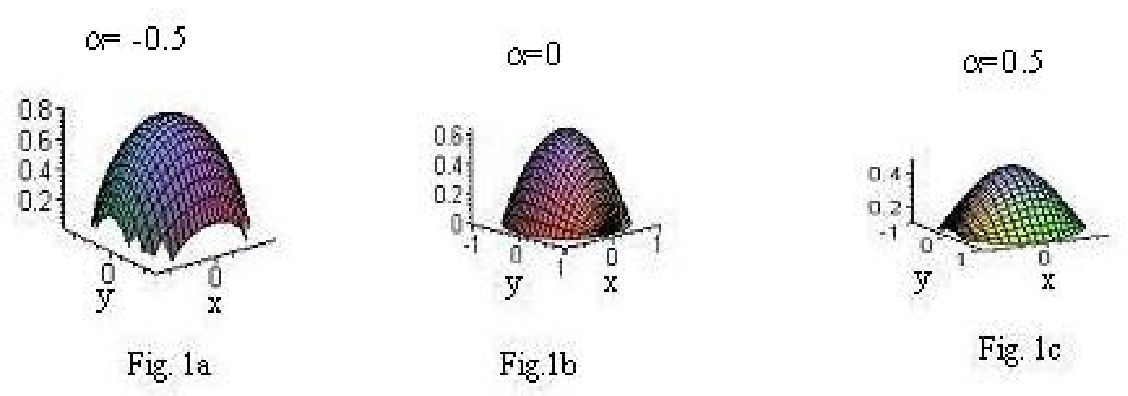}
\end{center}
For $\alpha>1$,
\[g_2(Y)=\frac{[a(\alpha-1)]^{\frac{q}{2}}\Gamma \left(\frac{1}{\alpha-1}\right)}
{\Gamma\left(\frac{1}{\alpha-1}-\frac{q}{2}\right)\pi ^{\frac{q}{2}}}
 [1+a(\alpha-1)(y_1^2+\cdots+y_q^2)]^{-\frac{1}{\alpha-1}},\alpha<1,
~\frac{1}{\alpha-1}-\frac{q}{2}>0,\alpha>1.\]
This is plotted for $q=2,a=1$, and for $\alpha=1.1,1.5,1.7.$
\begin{center}
\includegraphics{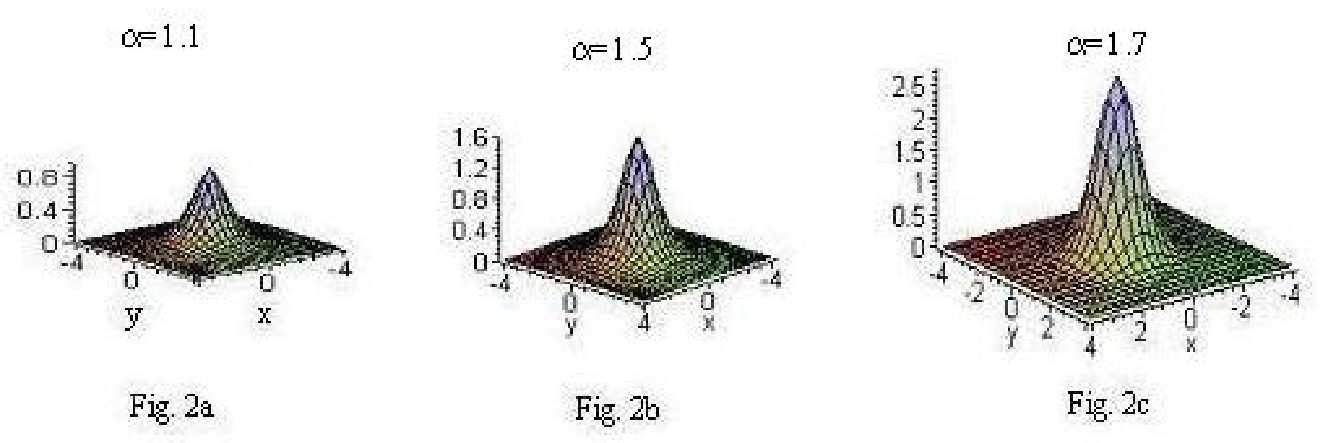}
\end{center}
For $\alpha\to 1$
\[g_3(Y)=\frac{a^{\frac{q}{2}}}{\pi^{\frac{q}{2}}}{\rm e}^{-a(y_1^2+\cdots+y_q^2)}.\]
\\
This is plotted for $a=1$.
\begin{center}
\includegraphics{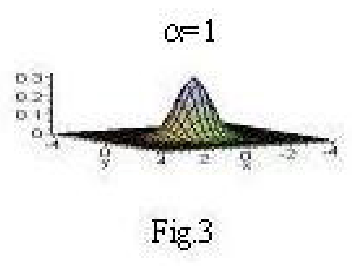}
\end{center}
 The nature of the pathway surface when $\alpha$ moves from -0.5 to 1 can be seen from Figures 1a-1c and Figure 3. The nature of the movement when $\alpha $ moves from 1 to 1.7 can be seen from Figure 3 and Figures 2a-2c. 

\section{Conclusions}

The multivariate Gaussian density and its central place in the procedure of maximizing a generalized entropic form of order $\alpha$ is the core result of this paper. It contributes to gain understanding of different entropic forms and how they relate to each other by using the parameter $\alpha$ (Mathai and Rathie [1], Masi [8]). This makes visible the pathway from type-1 beta, through Gaussian, to type-2 beta densities as they emerge depending on $\alpha$ and shows the relation to entropies of Boltzmann-Gibbs and Tsallis statistical mechanics (Hilhorst and Schehr [9], Vignat and Plastino [10]). While the generalized entropic form of order $\alpha$ may not have direct applications in statistical mechanics, it might be of interest to information theory and to a better understanding of attempts to unify entropic forms under either mathematical or physical principles. A graphical representation of the pathway is given in Figures 1, 2, and 3.
\medskip
\noindent

{\bf Acknowledgement} The authors would like to thank the Department of Science and Technology, Government of India, New Delhi, for the financial assistance for this work under project No. SR/S4/MS:287/05 which enabled this collaboration possible.\\
\vskip.5cm
\noindent

{\bf \large References}

\begin{enumerate}[{[1]}]
\item
A.M. Mathai and P.N. Rathie, {\it Basic Concepts in Information Theory and Statistics: Axiomatic Foundations and Applications}, Wiley Halsted, New York 1975.
\item 
A.M. Mathai and H.J. Haubold, Pathway model, superstatistics, Tsallis statistics, and a generalized measure of entropy, {\it Physica A} {\bf 375} (2007) 110-122.
\item
A.M. Mathai, A review of the recent developments on generalized complex matrix-variate Dirichlet integrals, in {\it Proceedings of the 7th International Conference of the Society for Special Functions and their Applications (SSFA)}, Pune, India, 21-23 February 2006, Ed. A.K. Agarwal, Published by SSFA, pp. 131-142.
\item
W. Feller, {\it An Introduction to Probability Theory and Its Applications}, Volume I, Third Edition, John Wiley and Sons, New York 1968.
\item
H.L. Swinney and C. Tsallis (Eds.), {\it Anomalous Distributions, Nonlinear Dynamics, and Nonextensivity}, {\it Physics D}, {\bf 193} (2004) 1-356.
\item
S. Abe and Y. Okamoto (Eds.), {\it Nonextensive Statistical Mechanics and Its Applications}, Springer, Heidelberg 2001.
\item
M. Gell-Mann and C. Tsallis (Eds.), {\it Nonextensive Entropy: Interdisciplinary Applications}, Oxford University Press, New York 2004.
\item
M. Masi, A step beyond Tsallis and Re'nyi entropies, {\it Physics Letters A}, {\bf 338} (2005) 217-224.
\item
H.J. Hilhorst and G. Schehr, A note on q-Gaussians and non-Gaussians in statistical mechanics, {\it Journal of Statistical Mechanics: Theory and Experiment}, 2007, P06003.
\item
C. Vignat and A. Plastino, Scale invariance and related properties of q-Gaussian systems, {\it Physics Letters A}, {\bf 365} (2007) 370-375.
\end{enumerate}
\end{document}